\documentclass{jpp}
\usepackage{graphicx}
\usepackage{natbib}

\usepackage[T1]{fontenc}
\usepackage{bm}
\usepackage{color}
\usepackage{graphicx}
\usepackage{amsmath}
\usepackage{amssymb}
\usepackage{amsfonts}
\usepackage{courier}
\usepackage{txfonts}
\usepackage{appendix}

\ifCUPmtlplainloaded \else
  \checkfont{eurm10}
  \iffontfound
    \IfFileExists{upmath.sty}
      {\typeout{^^JFound AMS Euler Roman fonts on the system,
                   using the 'upmath' package.^^J}%
       \usepackage{upmath}}
      {\typeout{^^JFound AMS Euler Roman fonts on the system, but you
                   dont seem to have the}%
       \typeout{'upmath' package installed. JPP.cls can take advantage
                 of these fonts, if you use 'upmath' package.^^J}%
      }
  \else
  \fi
\fi

\ifCUPmtlplainloaded \else
  \checkfont{msam10}
  \iffontfound
    \IfFileExists{amssymb.sty}
      {\typeout{^^JFound AMS Symbol fonts on the system, using the
                'amssymb' package.^^J}%
       \usepackage{amssymb}%

      }{}
  \fi
\fi

\ifCUPmtlplainloaded \else
  \IfFileExists{amsbsy.sty}
    {\typeout{^^JFound the 'amsbsy' package on the system, using it.^^J}%
     \usepackage{amsbsy}}
    {\providecommand\boldsymbol[1]{\mbox{\boldmath $##1$}}}
\fi

\newcommand{\unit}[1]{\ensuremath{\bm{\widehat{\mathrm{#1}}}}}
\usepackage[T1]{fontenc}
\usepackage{bm}
\usepackage{color}
\usepackage{graphicx}

\usepackage{ae}
\usepackage{amsmath}
\usepackage{amssymb}
\usepackage{amsfonts}
\usepackage{mathrsfs}
\usepackage{units}
\usepackage{natbib}
\usepackage[disable]{todonotes}
\newcommand{\g}[1]{\mbox{\boldmath $#1$}}

\newcommand{\be}{\begin{displaymath}}
\newcommand{\ee}{\end{displaymath}}
\newcommand{\bn}{\begin{equation}}
\newcommand{\en}{\end{equation}}

\usepackage{enumerate}

\newcommand{\nofrac}[2]{#1/#2}

\newcommand{\lnLc}{\ln\Lambda_{\rm c}}

\usepackage[T1]{fontenc}
\usepackage{bm}
\usepackage{color}
\usepackage{graphicx}
\usepackage[breaklinks=true]{hyperref}
\hypersetup{
  unicode=false,          % non-Latin characters in Acrobat’s bookmarks
  pdftoolbar=true,        % show Acrobat’s toolbar?
  pdfmenubar=true,        % show Acrobat’s menu?
  pdffitwindow=false,     % window fit to page when opened
  pdfstartview={FitH},    % fits the width of the page to the window
  pdfsubject={Subject},   % subject of the document
  pdfcreator={Creator},   % creator of the document
  pdfproducer={Producer}, % producer of the document
  pdfkeywords={keyword1} {key2} {key3}, % list of keywords
  pdfnewwindow=true,      % links in new PDF window
  colorlinks=true,        % false: boxed links; true: colored links
  linkcolor=blue,         % color of internal links (change box color with linkbordercolor)
  citecolor=blue,         % color of links to bibliography
  filecolor=blue,         % color of file links
  urlcolor=blue           % color of external links
}

\title[Current dynamics in elongated plasmas]{Effect of plasma elongation on current dynamics during tokamak disruptions} 
\author{T.~F\"ul\"op\aff{1} \corresp{\email{tunde@chalmers.se}}, P.~Helander\aff{2}, O.~Vallhagen\aff{1}, O.~Embr\'eus\aff{1}, L.~Hesslow\aff{1},
P.~Svensson\aff{1}, A.~J.~Creely\aff{3}, N.~T.~Howard\aff{4} and P.~Rodriguez-Fernandez\aff{4}}
\affiliation{\aff{1}Department of Physics, Chalmers University of Technology,
 SE-41296 G\"{o}teborg, Sweden
 \aff{2} Max Planck Institute for Plasma Physics,  Greifswald, Germany,  \aff{3} Commonwealth Fusion Systems, USA
 \aff{4} MIT Plasma Science and Fusion Center, Cambridge, Massachusetts 02139, USA}

\begin{document}
\maketitle
\begin{abstract}
Plasma terminating disruptions in tokamaks may result in relativistic runaway electron beams with potentially serious consequences for future devices with large plasma currents. In this paper we investigate the effect of plasma elongation on the coupled dynamics of runaway generation and resistive diffusion of the electric field. We find that elongated plasmas are less likely to produce large runaway currents, partly due to the lower induced electric fields associated with larger plasmas, and partly due to direct shaping effects, which mainly lead to a reduction in the runaway avalanche gain.
\end{abstract}

\maketitle

%%%%%%%%%%%%%%%%%%%%%%%%%%%%%%%%%%%%%%%%%%%%%%%%%%%%%%%%%%%

\section{Introduction}

Magnetic reconnection events in tokamaks often result in a sudden cooling of the plasma associated with an increase in the plasma resistivity, which in turn induces an electric field. If this electric field is larger than a certain critical electric field,  electrons in the tail of the bulk Maxwellian
distribution run away and acquire relativistic energies \citep{wilson1925,dreicer1959}.  These
 runaway electrons can multiply by producing additional runaway
electrons in close collisions with the thermal electrons, leading to
an exponential increase of the number of runaways -- an avalanche \citep{jayakumar1993}. 

Plasma-terminating disruptions in tokamaks often result in electric fields larger than the critical field. Since the avalanche production is exponentially sensitive to the initial plasma current, the problem with runaway generation is expected to be
more serious in future tokamaks with large plasma currents than in
present experiments \citep{RosenbluthPutvinski1997}. Uncontrolled loss of these high energy electrons
may cause serious damage to plasma facing components.

Most studies of runaway current dynamics have been performed assuming circular magnetic flux-surfaces, although both present and future devices often operate with elongated plasmas.  In particular future tokamaks with large plasma current, such as ITER  and SPARC  \citep{SPARC}, will have a significant elongation.  Experimental observations indicate that runaway beams are more easily produced in limiter or low-elongation discharges than in more elongated ones \citep{Izzo_2012,Hollmann_2013,Reux2015,Breizman_2019}. It is not clear if this is due to the elongation itself or to the difference of magnetic topology, i.e.~limited vs divertor configuration. MHD simulations show that differences in the MHD activity during the thermal quench phase produce better confinement of seed runaway electrons if the plasma is limited than if it is diverted  \citep{Izzo_2011}. Apart from these differences in MHD stability, there are also differences in the induced toroidal electric field and associated runaway current dynamics that depend on the plasma elongation directly, rather than indirectly via its effect on MHD stability.  

In this paper, we focus on such effects of elongation on the coupled
dynamics of runaway current generation and resistive electric field
diffusion. We derive an equation governing the evolution of the
toroidal electric field in general magnetic geometry and consider the
effect of elongation on the current dynamics in the case of an
axisymmetric magnetic field with elliptical flux-surfaces in the large
aspect ratio limit. We show that the effect of elongation is to reduce
both the maximum electric field, leading to lower runaway generation
rate, as well as the potential runaway avalanche multiplication. We
then illustrate the effect of elongation in simulated
{idealized} disruptions, {with a
pre-described exponentially decaying temperature evolution}. 
\section{Evolution of toroidal electric field}
The magnetic field in general toroidal geometry can be written as
\begin{align}
\g B =\nabla \psi \times  \nabla \theta+\nabla \varphi\times\nabla \chi
\label{Bfieldgeneral}
\end{align}
where $\psi$ and $\chi$ are the toroidal and poloidal fluxes divided by $2  \pi$, and $\theta$ and $\varphi$ are magnetic coordinates corresponding to poloidal and toroidal angles, respectively.  We are interested in the evolution of the flux-surface average of the toroidal component of the induced electric field $\g E=-\partial \g A/\partial t-\nabla\phi$, where $\g A$ and $\phi$ are the vector potential and the electrostatic potential, respectively. Following Appendix A of \citet{Boozer_2018}, we find that the electric field can be written as
\begin{align}
    \g E=\left(\frac{\partial \chi}{\partial t}\right)_\psi \nabla \varphi -\g u\times \g B -\nabla (s+\phi)
    \label{efield}
\end{align}
with $\g u =(\partial \g r/\partial t)_{\psi,\theta,\varphi}$ the velocity of the canonical coordinates $(\psi,\theta,\varphi)$ through space, $s=\psi\dot{\theta}-\chi \dot{\varphi}$, and the dot denotes a time derivative taken at constant $\g r$. The projection along $\g B$ of the two last terms in equation~(\ref{efield}) vanish upon flux-surface averaging, thus the evolution of the parallel electric field in an arbitrarily shaped tokamak or stellarator is given by 
\begin{align}
\langle \g E \cdot \g B\rangle = \left(\frac{\partial \chi}{\partial t}\right)_\psi \langle \g B\cdot \nabla \varphi\rangle,
\label{EB}
\end{align}
where the flux surface average is defined as 
$$\langle\zeta\rangle ({\chi})= 
\int \zeta ({\chi},\theta, \varphi) dV^\prime \bigg\slash \int dV'
 $$
where the integrals are taken over the volume between two neighbouring flux surfaces and the volume element has been written as $dV' = {({\g B} \cdot \nabla \theta)^{-1}d\psi} d\theta d \varphi$.

In an {\em axisymmetric} magnetic field, we can write \citep{helander}
\begin{align}
\g B =F(\chi)  \nabla \varphi+\nabla \varphi\times\nabla \chi,
\label{Bfieldaxi}
\end{align}
which simplifies the toroidal electric field to
\begin{align}
\langle \g E\cdot \g B\rangle =F\,\langle R^{-2}\rangle \left(\frac{\partial \chi}{\partial t}\right)_\psi,
\label{EBaxi}
\end{align}
where $R = |\nabla \varphi|^{-1}$ denotes the major radius. 

If we neglect the contribution of the plasma pressure in the equilibrium equation $\g j\times \g B=\nabla p$ the current has the form $\mu_0\g j =K(\chi,t)\g B$, so that, using Amp\`ere's law $\mu_0\g j =\nabla\times \g B$,
we find \begin{align}\mu_0 \g j \cdot\nabla \varphi = KF/R^2= (\nabla\times \g B)\cdot \nabla\varphi=\nabla \cdot (\g B\times\nabla\varphi)=\nabla \cdot (R^{-2}\nabla \chi).
\label{jaxi}
\end{align}
Upon flux-surface averaging we then have
\begin{align}
K=\frac{1}{F\langle R^{-2}\rangle}\left\langle\nabla \cdot\left(\frac{\nabla\chi}{R^2}\right)\right\rangle=\frac{1}{F\langle R^{-2}\rangle{V^\prime}}\frac{\partial}{\partial r} V^\prime\left\langle \left|\frac{\nabla r}{R}\right|^2 \right\rangle\frac{\partial \chi}{\partial r},
\label{K}
\end{align}
where we have introduced a coordinate $r$ labelling flux surfaces, the volume $V(r)$ enclosed by such surfaces and the prime denotes a derivative with respect to $r$.

We now specialise further to the limit of large aspect ratio, in which the current density $j_\|/  R={\g j} \cdot \nabla \varphi $ and the electric field $E \simeq R^{-1} \partial \chi / \partial t$ are approximately constant over each magnetic surface. Taking a time derivative of equation~(\ref{jaxi}) and using equation~(\ref{EBaxi}), we find that these quantities satisfy the following equation: 
\begin{align}
  \mu_0 \frac{\partial j_\parallel}{\partial t} = \frac{1}{ V^\prime}\frac{\partial}{\partial r}V^\prime \langle |\nabla r|^2\rangle \frac{\partial E}{\partial r}. 
\label{LAS1}
\end{align}
If the magnetic surfaces have elliptical cross section with elongation $\kappa$ (which can depend on radius) and we let $r$ denote the radius along the minor axis, then 
$
V(r)=2\pi^2 R \kappa r^2
$
so that $V^\prime (r)= 2 \pi^2 R (\kappa^\prime r^2+ 2 \kappa r)$. Since $r^2=x^2+y^2/\kappa^2$ we have $r\nabla r=x\nabla x+y\nabla y/\kappa^2-(y^2\kappa^\prime/\kappa^3) \nabla r$, which in terms of the polar angle $\theta$ ($\tan{\theta}= \kappa^{-1} y/x$) takes the form 
$$
|\nabla r|^2=\frac{\cos^2{\theta}+\kappa^{-2}\sin^2{\theta}}{\left(1+\frac{\kappa^\prime r}{\kappa}\sin^2{\theta}\right)^2}.
$$ 
We thus have, with the Jacobian {$ \left| \partial(x,y)/\partial (r,\theta)\right|= \kappa r +  \kappa' r^2 \sin^2 \theta$},
\begin{align}
    \hspace{-3mm} \langle  |\nabla r|^2\rangle
    =\frac{\kappa r}{\pi(2\kappa r+\kappa^\prime r^2)}\int_0^{2\pi}\frac{1-(1-\kappa^{-2})\sin^2{\theta}}{1+\frac{\kappa^\prime r}{\kappa}\sin^2{\theta}}d\theta 
    = \frac{\kappa^{-2} + \sqrt{1+ \frac{\kappa' r}{\kappa}}}{\left(1+\frac{\kappa' r}{2\kappa}\right)\left(1+\sqrt{1+\frac{\kappa' r}{\kappa}}\right)\sqrt{1+\frac{\kappa' r}{\kappa}}},
    \label{generalnablar}
\end{align}
which can be substituted into equation~(\ref{LAS1}) to give an equation for the current density evolution in the case of an arbitrary elongation. {As most of the runaway generation occurs close to the magnetic axis, we  can neglect finite aspect ratio effects here, which would give corrections only of order $(r/R)^2$. }

If the elongation is constant, this equation simplifies to
\begin{align}
\mu_0 \frac{\partial j_\parallel}{\partial t}=\frac{1+\kappa^{-2}}{2}\frac{1}{r}\frac{\partial }{\partial r} r\frac{\partial E}{\partial r}.
\label{GO}
\end{align}
Hence it is apparent that elongation affects the resistive diffusion of the electric field. Moreover, at fixed total plasma current and minor semiradius of the elliptical plasma cross section, the current density is inversely proportional to $\kappa$. The induced electric field before and immediately after the thermal quench depends similarly on $\kappa$. The  Dreicer runaway production rate \citep{dreicer1959,connor}, which is exponentially sensitive to the electric field, can therefore be reduced significantly by finite elongation.

\section{Runaway generation and evolution of plasma current}
The evolution of the current density is governed by a balance between the generation of the runaway electrons and the resistive diffusion of the electric field accelerating them \citep{Eriksson2004}. In the cylindrical case,  a 1D model of these processes is numerically evolved by the {\sc go} code \citep{Smith2006GO}, in which the current density is assumed to be the sum of the Ohmic and runaway current densities. The runaways are assumed to travel at the speed of light, so that  $j_\| =\sigma E+ecn_\text{RE}$, where $n_\text{RE}$ is the number density of runaways. {\sc go} has been used extensively for evaluating pellet- and gas-injection scenarios in JET and ITER-like plasmas \citep{gal,Feher,HollmannDMS,Hesslow_2019}, and for interpretative modelling of experiments, e.g.~the effect of the wall material on RE beam formation in the JET tokamak \citep{Papp_2013}.

As the resistivity increases in connection with the thermal quench, an electric field is induced, which gives rise to a runaway seed population by velocity space diffusion into the runaway region due to small angle collisions \citep{dreicer1959}. To determine the Dreicer runaway growth rate accurately, we use a neural network\footnote{The neural network is available at \href{https://github.com/unnerfelt/dreicer-nn}{https://github.com/unnerfelt/dreicer-nn}} \citep{Hesslowupcoming}
trained on a large number of kinetic simulations by the {\sc code} kinetic equation solver \citep{CODEPaper2014}. In the case of fully ionized plasmas and constant Coulomb logarithm, the primary runaway growth rate given by the neural network agrees with the analytical formulas for the runaway growth rate of \citet{connor}. However, due to the velocity-dependence of the Coulomb logarithm, in certain regions of parameter space, in particular for low temperatures and electric fields, the growth rate can significantly differ from the analytical formulas, even in fully ionized plasmas \citep{Hesslowupcoming}. This leads to substantial changes in the final runaway current, as we will see in the next section.

The primary effect of the Dreicer mechanism is to generate a ``seed'' of runaways which is amplified by the avalanche, but there are also other ways in which seeding occurs. For instance, tritium decay produces a seed which can be modelled as \citep{MartinSolis2017}
\begin{equation}
    \left(\frac{\partial n_\mathrm{RE}}{\partial t}\right)^\mathrm{tritium}=\ln{(2)}\frac{n_\mathrm{T}}{\tau_\mathrm{T}}f\left(W_\mathrm{crit}\right),
\end{equation}
where $n_\mathrm{T}$ is the tritium density, $\tau_\mathrm{T}\approx4500$ days is the half life of tritium, and $f(W_\mathrm{crit})$ is the fraction of the electron spectrum of the tritium decay above the critical runaway energy $W_\mathrm{crit}$.   If we consider the emitted electron as a free particle, i.e.~neglect the effect of the Coulomb interaction between the electron and the tritium nucleus on the spectrum, the tritium energy spectrum can be shown to fulfill \citep{subatomic} $
    I(W)\propto(Q-W)^2\sqrt{W}$, where $Q=18.6\,\rm keV$. The fraction of the electron spectrum from tritium decay that lies within the runaway region can then be calculated analytically as
\begin{equation}
    f(W_\mathrm{crit})=\frac{\int_{W_\mathrm{crit}}^Q I(W)\mathrm{d}W}{\int_\text{0}^Q I(W)\mathrm{d}W}=1-\frac{35}{8}\left(\frac{W_\mathrm{crit}}{Q}\right)^{3/2}+\frac{21}{4}\left(\frac{W_\mathrm{crit}}{Q}\right)^{5/2}-\frac{15}{8}\left(\frac{W_\mathrm{crit}}{Q}\right)^{7/2},
    \label{eq:frac_RE_beta}
\end{equation}
where $W_\mathrm{crit}$ is the critical runaway energy.

The seed runaways are amplified due to close collisions. In fully ionized plasmas, the avalanche growth rate is given by \citep{RosenbluthPutvinski1997}
\begin{eqnarray}
 \left(\frac{\partial n_\text{RE}}{\partial t}\right)^\text{aval}= \frac{e(E_\parallel-E_c){n_\text{RE}}}{m_{\rm e} c \lnLc} {\sqrt{\frac{\pi \varphi_\epsilon}{3\textcolor{black}{(Z_{\rm
        eff}+5)}}} \left( 1-
    \frac{E_{c}}{E_\parallel}+\frac{4\pi (Z_{\rm eff}+1)^2}{3\varphi_\epsilon (Z_{\rm
        eff}+5)(E_\parallel^2/E_{c}^2+4/\varphi_\epsilon^2-1)} \right)^{-1/2}},
\label{aval}
\end{eqnarray}
where $E_{\rm c}\,{=}\,m_{\rm e} c/(e \tau_{\rm c})$ is the
Connor--Hastie critical electric field,
$\tau_\text{c}\,{=}\,\nofrac{4 \pi \epsilon_0^2 m_{\rm e}^2
  c^3}{(n_{\rm e} e^4 \lnLc)}$ is the relativistic collision time, 
$ \lnLc \,{\approx}\,14.6+0.5 \ln (T_{\rm eV}/n_{\rm e20})$ is the
Coulomb logarithm for relativistic electrons, with $T_{\rm eV}$ the
electron temperature in electronvolt, $n_{\rm e20}$ the density of
the background electrons in units of $\unit[10^{20}]{m^{-3}}$, 
{$\varphi_\epsilon=(1+1.46\epsilon^{1/2}+1.72
  \epsilon)^{-1}$ is the neoclassical factor and $\epsilon=r/R$ denotes the inverse aspect ratio}. {The effect of elongation on the neoclassical factor $\varphi_\epsilon$ is negligible. } 
Note
that, in the presence of partially ionized impurities, the avalanche
growth rate will no longer be directly proportional to the electric
field, and the multiplication factor will instead become sensitive to
the details of the electric field evolution \citep{martinsolis1,Hesslow_2019}, but
in fully ionized plasmas which we consider here, these effects can be
ignored. 
\section{Dependence of avalanche multiplication on plasma elongation}
When primary runaway generation is negligible, the plasma current evolution is governed by the diffusion of the electric field and runaway avalanche multiplication according to equations~(\ref{LAS1}) and (\ref{aval}), 
\begin{align}
    \mu_0 \frac{\partial j_\parallel}{\partial t}=\frac{1}{V^\prime}\frac{\partial}{\partial r}V^\prime \langle |\nabla r|^2\rangle \frac{\partial E_\parallel}{\partial r} \label{diffusion}\\
    \frac{\partial \ln{n_\text{RE}}}{\partial t}=\frac{E_\parallel}{E_\text{c} \tau_\text{a}} 
    \label{avalanche}
\end{align}
where $\tau_\text{a}=\tau_ \text{c}\lnLc\sqrt{5+Z_{\rm eff}}$ and we have assumed $E_\parallel \gg E_\text{c}$. 

If the density, effective charge and the Coulomb logarithm are constant in time, we can integrate the diffusion equation (\ref{diffusion}) from $t=0$ to $t=\infty$ when the entire current is carried by runaways:
\begin{align}
    -\mu_0\left[j_\text{0}-j_\text{RE}\right]=\frac{1}{ V^\prime}\frac{\partial}{\partial r}V^\prime \langle |\nabla r|^2\rangle \frac{\partial}{\partial r}\left[E_\text{c} \tau_\text{a} \ln{\frac{j_\text{RE}(r)}{ec n_\text{seed}(r)}}\right].
\end{align}

 To find the maximum possible avalanche multiplication in the trace runaway limit, $j_\text{RE}\ll j_\text{0}$,  we integrate in radius from $0$ to $r$ and obtain
 \begin{align}
    2\pi R \mu_0 I_\text{0}(r)=- V^\prime(r) \langle  |\nabla r|^2 \rangle   \frac{\partial}{\partial r}\left[E_\text{c} \tau_\text{a} \ln{\frac{j_\text{RE}(r)}{e c n_\text{seed}(r)}} \right],
 \end{align}
 where $I_\text{0} = (2\pi R)^{-1}\int_0^r dr^\prime \, V^\prime(r^\prime) j(r^\prime)$ denotes the total initial current enclosed by a flux surface of minor radius $r$.
Integrating again from $r$ to  $a$, where we assume a perfectly conducting wall, we obtain the avalanche multiplication factor that accounts for the radial diffusion of the electric field,
\begin{align}
    \ln{\frac{j_\text{RE}(r)}{e c n_\text{seed}(r)}} = \frac{\mu_0}{E_\text{c}(r) \tau_\text{a}(r)}\int_r^a \frac{2\pi R I_\text{0}(r^\prime)\, dr^\prime}{V^\prime (r^{\prime}) \langle  |\nabla r|^2 \rangle (r^{\prime})}.
    \label{jfrac}
\end{align}
In a geometry with constant elongation,   
$\langle  |\nabla r|^2\rangle=(1+\kappa^{-2})/2 $ and $V^\prime (r)= 4 \pi^2 \kappa r R$, and assuming that the plasma current $I$ is independent of the elongation, equation (\ref{jfrac}) gives
\begin{align}
    \ln{\frac{j_\text{RE}}{ec n_\text{seed}}} =\frac{\mu_0 G(\kappa)}{2\pi E_\text{c} \tau_a}\int_r^a \frac{dr^{\prime}}{r^\prime}I_\text{0}(r^\prime)
\end{align}
where $G(\kappa)=2/(\kappa+\kappa^{-1})$. Thus the avalanche generated runaway current will be reduced by a factor of $\exp{\left[N_\mathrm{exp}^{\kappa=1}\left(1-G(\kappa)\right)\right]}$, where $N_\mathrm{exp}^{\kappa=1}$ is the number of exponentiations for $\kappa=1$. In the case of $\kappa=1.45$ used in the examples in the next section, we have $G(1.45)=0.93$. With a representative value of $N_\mathrm{exp}\approx 50$, for ITER-like parameters in a fully ionized plasma \citep{RosenbluthPutvinski1997}, this would correspond to a reduction factor of about 26.
However, this argument applies only in the case when the final runaway current is much smaller than the initial one. For higher runaway currents, the effect of the runaway current on the electric field will cause the runaway current to saturate and hence the number of exponentiations can be much lower.

\section{Numerical results for high current devices}
To illustrate the effect of elongation, in the following, we present numerical solutions of equation~(\ref{GO}), with the runaway growth rate given by the sum of the primary (Dreicer+tritium decay) and avalanche growth rates, for parameters characteristic of a SPARC V0 and an ITER discharge.  We take the temperature to decay exponentially as $T_\text{e}(t, x)=T_\text{f}(x)+\left[T_\text{0}(x)-T_\text{f}(x)\right] e^{-t/t_\text{0}}$, where $T_\text{0}$ and $T_\text{f}$ are the initial and final electron temperatures, respectively, and $t_\text{0}$ is the thermal quench (TQ) time. For simplicity, we consider pure plasmas and neglect hot-tail generation and runaways produced by Compton scattering of $\gamma$-rays due to the activated wall in the nuclear phase of operation.  In view of all these assumptions, the results can only be used as an illustration of the effect of elongation, and not to draw conclusions on the final runaway current in any future tokamak.

\begin{figure}
     \centering
     \includegraphics[width=0.9\textwidth]{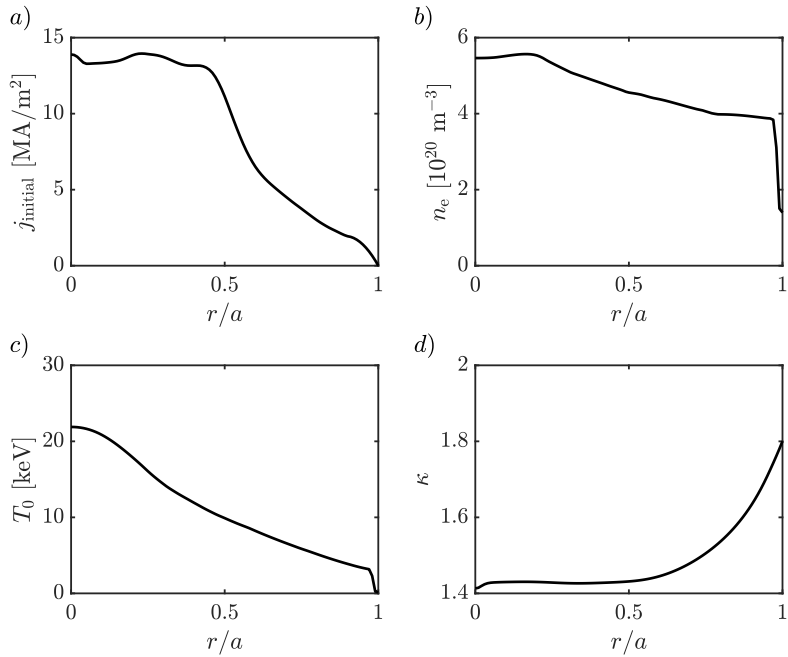} 
     \caption{Radial profiles of plasma parameters for SPARC V0. (a) Initial current density. (b) Electron density. (c) Electron temperature. (d) Elongation.}
     \label{profiles}
 \end{figure}
For SPARC V0, the initial plasma current is $I_\text{0}=7.5\,\rm MA$, major radius $R=1.65\,\rm m$ and minor radius $a=0.5\,\rm m$. The initial current density, electron density, temperature and elongation profiles are shown in figure~\ref{profiles}. Profiles were obtained from predictive modelling using the {\sc transp} code \citep{trams}, which takes into account sources, sinks and transport projected to be present in SPARC V0.  In the simulations presented below we keep the total plasma current constant when changing the elongation, which means that the local current density is rescaled correspondingly. 

After a disruption, the final temperature profile is usually flatter than the initial one, and we will assume it to be constant $T_\text{f}=20 \,\rm eV$. The final density is normally larger than before the disruption, due to an influx of impurities from the wall or intentional injection of gas to mitigate the effects of the disruption. Here, for simplicity we take the density to be constant in time with the same radial profile as the initial density. 

\begin{figure}
     \centering
     \includegraphics[width=\textwidth]{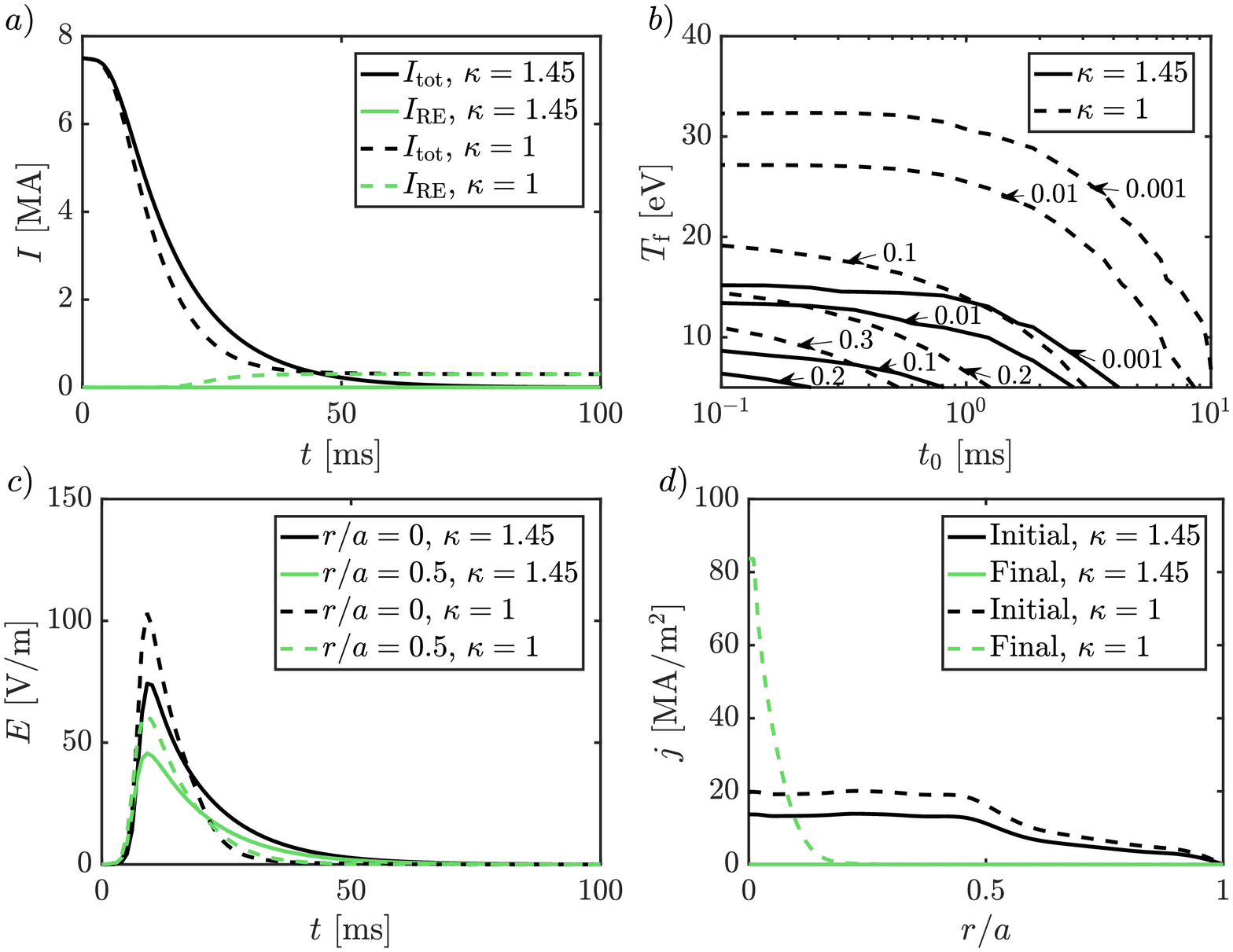}
     \caption{Plasma current and electric field evolution in a simulated SPARC V0 thermal quench. (a) Total plasma current as function of time. Dotted lines correspond to circular plasma ($\kappa=1$), and dashed lines are for ($\kappa=1.45$). (b) Contour plot of the current conversion $I_\text{RE}/I_\text{tot}$ as function of TQ time $T_\text{0}$ and final temperature $T_{f}$ for $\kappa=1$ (dashed) and $\kappa=1.45$ (solid). (c,d) Electric field and current density evolution for circular (dashed) and elongated (solid) plasmas. The parameters are $t_\text{0}=1\,\rm ms$ and $T_\text{f}=20\,\rm eV$, except in (b), where they are varied.}
     \label{fig:current}
 \end{figure}

Figure \ref{fig:current} shows the electric field and current evolution in a simulated SPARC V0 thermal quench, for both elongated and circular plasma shapes.  As figure~\ref{profiles}d shows, the elongation varies radially, but our simulations show that what matters for runaway generation is the value of the elongation in the central part of the plasma. For SPARC V0 we find that the runaway current evolution is the same when we take into account the radial dependence of the elongation as when we take the value $\kappa=1.45$. As the difference to the radially varying case is insignificant,   figure~\ref{fig:current} only shows the electric field and current evolution for a constant value of $\kappa=1.45$. 

Figure~\ref{fig:current}a shows that, if the plasma is elongated,  only a negligible part  of the initial plasma current is converted to a runaway current, compared to the circular case.  Figure \ref{fig:current}b shows the current conversion as a function of TQ time and final temperature in the elongated case. Clearly, runaway production is not significant unless the cooling is extremely rapid and the final temperature reaches less than a few eV.

The simulations show that the main reason for the reduction in the current conversion in elongated plasmas is the lower maximum electric field compared to circular plasmas, as shown in figure~\ref{fig:current}c. The change in electric field  significantly affects the Dreicer generation. The avalanche multiplication factor is also reduced for $\kappa>1$, as shown in the previous section. The maximum of the electric field and consequently the runaway production is mainly on-axis, as shown in figure~\ref{fig:current}d. 

Including tritium seed generation leads to a runaway current conversion of $4.4\%$ in the cylindrical case and $1.2\%$ in the elongated case for TQ time $t_\text{0}=1\,\rm ms$ and final temperature $T_\text{f}=20\,\rm eV$.  The corresponding numbers without tritium are  $4.1\%$ in the cylindrical case and $10^{-6}$ in the elongated case. For $T_\text{f}=20\,\rm eV$, Dreicer and tritium runaway seeds are of the same order of magnitude in the cylindrical case, but the tritium seed dominates in the elongated case.

For the ITER scenario we consider, the initial plasma current is $I_\text{0}=15\,\rm MA$ and the  major and minor radii are $R=6\,\rm m$ and $a=2\,\rm m$, respectively. The pre-disruption average temperature is $\langle T_\text{e}\rangle = 10\,\rm keV$ and density $\langle n_\text{e}\rangle =10^{20}\,\rm m^{-3}$. The temperature profile is taken as $T_\text{e}=T_{0} \left[1-(r/a)^2\right]$, with $T_{0}=2\langle T_\text{e}\rangle$, and the electron density $n_\text{e}$ profile is assumed to be flat.  The initial current density profile is assumed to be $j(r)=j_\text{0} \left[1-(r/a)^{0.41}\right]$, corresponding to an internal inductance of $l_\text{i}=0.7$. $j_\text{0}$ is a normalization parameter chosen so that the total plasma current integrates to $I_p$. For $\kappa=1$ it is $j_\text{0}=1.69 \,\rm MA/m^2$ and for $\kappa=1.45$, $j_\text{0}=1.16\, \rm MA/m^2$. These parameters and initial profiles are similar to the ones used by \citet{MartinSolis2017}, except for the effect of elongation which was not considered there. 

\begin{figure}
     \centering
     \includegraphics[width=\textwidth]{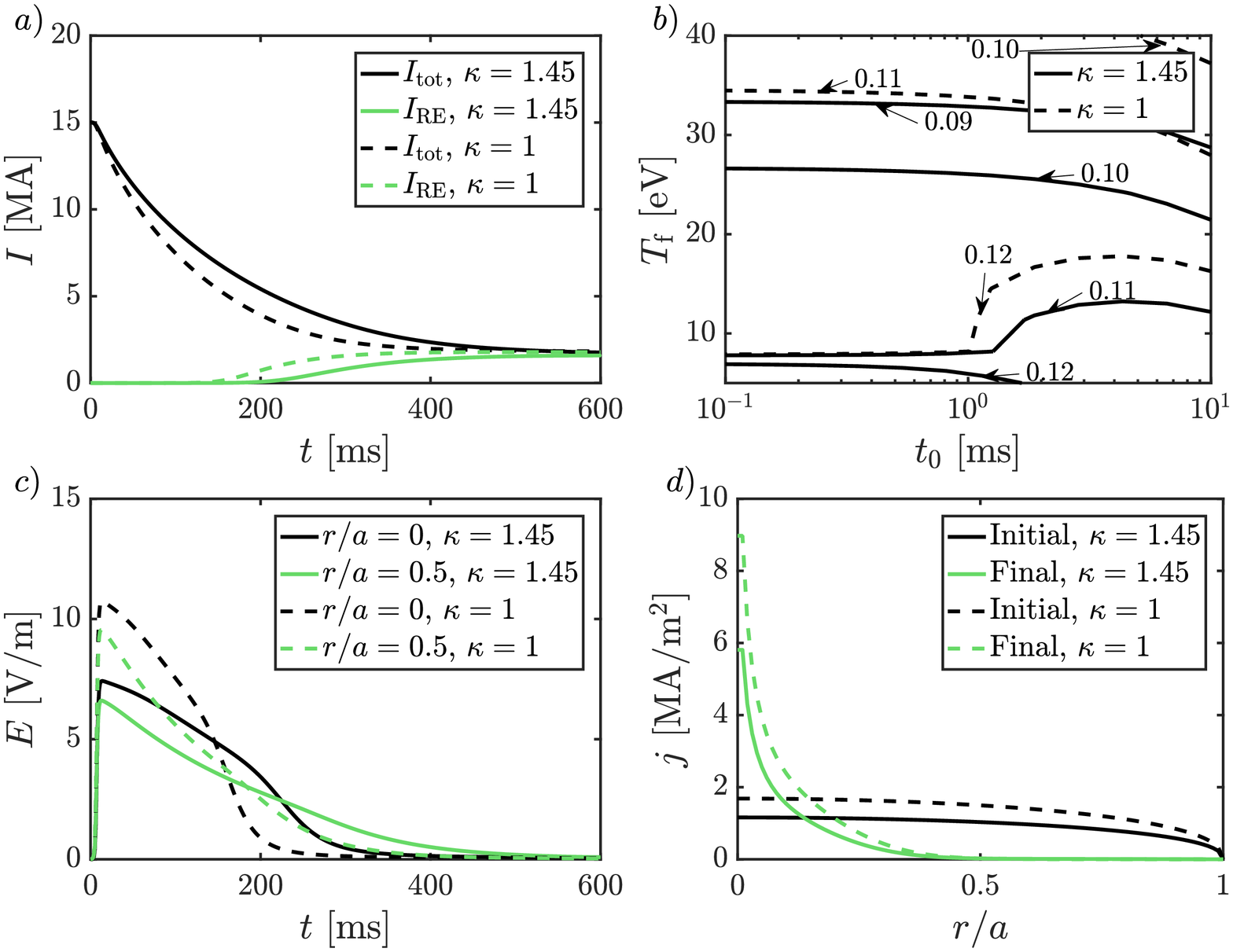}
     \caption{Plasma current and electric field evolution in a simulated ITER thermal quench. (a) Total plasma current as function of time for circular (dashed) and elongated (solid) plasmas. (b) Contour plot of the current conversion $I_\text{RE}/I_\text{tot}$  as function of TQ time $t_\text{0}$ and final temperature $T_{f}$ for $\kappa=1.45$. (c,d) Electric field and current density evolution. The parameters are $t_\text{0}=1\,\rm ms$ and $T_{f}=20\,\rm eV$ except in (b) where they vary.}
     \label{fig:currentITER}
 \end{figure}

Figure \ref{fig:currentITER} shows the current and electric field evolution in a simulated ITER disruption, comparing a circular and an elongated plasma, with constant elongation $\kappa=1.45$. In the considered case, the maximum electric field is much lower in ITER than in SPARC. Therefore, the Dreicer part of the runaway seed is negligibly small, and the tritium seed dominates with orders of magnitude. This was noted also in previous work, e.g.~by \cite{MartinSolis2017}. 

The tritium seed generation is primarily determined by the time interval during which the electric field is high enough for the critical energy for runaway generation $W_{crit}$ to be lower than the maximum energy of the emitted electrons during tritium decay $Q=18.6\,\rm keV$. This interval increases with elongation, and consequently the tritium seed slightly increases with $\kappa$. However, the final runaway current is still reduced, but only marginally. The reason for the reduction is that the avalanche multiplication is weaker in the elongated case.
The sensitivity of the current conversion  to the TQ time and final temperature is shown in figure~\ref{fig:currentITER}b. 

 We do not consider the effect of shaping on the MHD stability of the
 discharge, which might give rise to radial transport of energetic
 runaways, or any other loss processes.  We have also ignored several
 processes that would lead to higher runaway currents. Perhaps one of
 the most important of these is hot-tail generation of runaways which
 is expected to be significant in large tokamak disruptions
 \citep{Chiu1998,helanderhottail,MartinSolis2017,Aleynikov2017}.
 Hot-tail generation is very sensitive to the details of the cooling
 process following the magnetic reconnection, which is not well
 understood \citep{BreizmanAleynikov2017Review}. In addition, as the
 hot-tail runaways are produced in the early phase of the TQ, their
 transport is likely to be significantly affected by the high level of
 magnetic fluctuations following the magnetic reconnection. Therefore
 it is difficult to obtain accurate predictions regarding the effect
 of hot-tail generation.

 {An additional limitation of
   the analysis above is the assumption of a pre-described
   exponentially decaying temperature evolution, with a flat final
   temperature profile. In a more realistic scenario, the temperature is
   determined by a balance between Ohmic heating and impurity
   radiation, and it will have an important effect on the evolving
   current density profile. To investigate whether or not the effect
   of elongation changes if we use a different temperature profile, we
   have simulated a case where the radial temperature profile is kept
   constant (i.e.~the same as the initial). We find that the effect of
   elongation is not affected by this change. For ITER, the change in the final runaway current is negligible;  the
only effect is that the current quench becomes somewhat shorter and
the electric field larger, but during a shorter time, leading to the
same final runaway current. For SPARC, where Dreicer generation
dominates, the difference in the final current is somewhat larger but
the trend of how the elongation reduces the current is the same. }

 Currently envisaged disruption mitigation methods involve injection of massive amounts of material, and that will also change the current dynamics substantially, and may lead to higher runaway currents \citep{Hesslow_2019}.  On the other hand, in connection with material injection, the injected density required to raise the critical field $E_c$ for runaway generation (or the threshold field for tritium seed generation) above the maximum induced field will be lower by a factor of $\kappa$ in elongated plasmas.

\section{Conclusions}
We show that elongated plasmas are less prone to runaway electron generation in tokamak disruptions. Since the current density is approximately inversely proportional to the vertical elongation $\kappa$, the maximum induced electric field is reduced by a similar factor, which has a significant effect on the primary runaway generation, which is exponentially sensitive to the electric field. In addition, shaping reduces the maximum avalanche gain by a factor of $2/(\kappa+\kappa^{-1})$.  Numerical solution of the coupled equations of runaway generation and resistive diffusion of electric field in simulated disruptions in high-current devices show that the final runaway current is expected to be reduced considerably in tokamaks where the primary runaway generation is dominated by the Dreicer process. When the primary generation is dominated by other processes, such as tritium decay, we expect the elongation to cause only a marginal reduction of the final runaway current.
 \bibliographystyle{jpp}
\section*{Acknowledgements} The authors are grateful to S.~Newton, M.~Hoppe, L.~Unnerfelt,  A.~Tinguely and G.~Papp for fruitful discussions. This project has received funding from the European Research Council (ERC) under the European Union's Horizon 2020 research and innovation programme under grant agreement No 647121. The work was also supported by the Swedish Research Council (Dnr.~2018-03911), the EUROfusion - Theory and Advanced Simulation Coordination (E-TASC) and was carried out within the framework of the EUROfusion Consortium and has received funding from the Euratom research and training programme 2014-2018 and 2019-2020 under grant agreement No 633053. The views and opinions expressed herein do not necessarily reflect those of the European Commission.
\bibliography{references} % references.bib

\begin{thebibliography}{30}
\expandafter\ifx\csname natexlab\endcsname\relax\def\natexlab#1{#1}\fi
\def\au#1{#1} \def\ed#1{#1} \def\yr#1{#1}\def\at#1{#1}\def\jt#1{\textit{#1}}
  \def\bt#1{#1}\def\bvol#1{\textbf{#1}} \def\vol#1{#1} \def\pg#1{#1}
  \def\publ#1{#1}\def\arxiv#1{#1}\def\org#1{#1}\def\st#1{\textit{#1}}

\bibitem[Aleynikov \& Breizman(2017)]{Aleynikov2017}
{\sc \au{Aleynikov, P.} \& \au{Breizman, B.~N.}} \yr{2017}  \at{Generation of
  runaway electrons during the thermal quench in tokamaks}.  \jt{Nuclear
  Fusion}  \bvol{57}~(4),  \pg{046009}.

\bibitem[Boozer(2018)]{Boozer_2018}
{\sc \au{Boozer, A.~H.}} \yr{2018}  \at{Pivotal issues on relativistic
  electrons in {ITER}}.  \jt{Nuclear Fusion}  \bvol{58}~(3),  \pg{036006}.

\bibitem[Breizman \& Aleynikov(2017)]{BreizmanAleynikov2017Review}
{\sc \au{Breizman, B.} \& \au{Aleynikov, P.}} \yr{2017}  \at{Kinetics of
  relativistic runaway electrons}.  \jt{Nuclear Fusion}  \bvol{57}~(12),
  \pg{125002}.

\bibitem[Breizman {\em et~al.\/}(2019)Breizman, Aleynikov, Hollmann \&
  Lehnen]{Breizman_2019}
{\sc \au{Breizman, B.~N.}, \au{Aleynikov, P.}, \au{Hollmann, E.~M.} \&
  \au{Lehnen, M.}} \yr{2019}  \at{Physics of runaway electrons in tokamaks}.
  \jt{Nuclear Fusion}  \bvol{59}~(8),  \pg{083001}.

\bibitem[Breslau {\em et~al.\/}(2018)Breslau, Gorelenkova, Poli, Sachdev \&
  Yuan]{trams}
{\sc \au{Breslau, J.}, \au{Gorelenkova, M.}, \au{Poli, F.}, \au{Sachdev, J.} \&
  \au{Yuan, X.}} \yr{2018} {TRANSP}. [Computer Software]
  \url{https://doi.org/10.11578/dc.20180627.4}.

\bibitem[Chiu {\em et~al.\/}(1998)Chiu, Rosenbluth, Harvey \& Chan]{Chiu1998}
{\sc \au{Chiu, S.}, \au{Rosenbluth, M.}, \au{Harvey, R.} \& \au{Chan, V.}}
  \yr{1998}  \at{Fokker-planck simulations mylb of knock-on electron runaway
  avalanche and bursts in tokamaks}.  \jt{Nuclear Fusion}  \bvol{38}~(11),
  \pg{1711--1721}.

\bibitem[Connor \& Hastie(1975)]{connor}
{\sc \au{Connor, J.} \& \au{Hastie, R.}} \yr{1975}  \at{Relativistic
  limitations on runaway electrons}.  \jt{Nuclear Fusion}  \bvol{15},
  \pg{415}.

\bibitem[Dreicer(1959)]{dreicer1959}
{\sc \au{Dreicer, H.}} \yr{1959}  \at{Electron and ion runaway in a fully
  ionized gas. {I}}.  \jt{Phys. Rev.}  \bvol{115},  \pg{238}.

\bibitem[Eriksson {\em et~al.\/}(2004)Eriksson, Helander, Andersson, Anderson
  \& Lisak]{Eriksson2004}
{\sc \au{Eriksson, L.-G.}, \au{Helander, P.}, \au{Andersson, F.}, \au{Anderson,
  D.} \& \au{Lisak, M.}} \yr{2004}  \at{Current dynamics during disruptions in
  large tokamaks}.  \jt{Phys. Rev. Lett.}  \bvol{92},  \pg{205004}.

\bibitem[Feh\'er {\em et~al.\/}(2011)Feh\'er, Smith, F\"ul\"op \& G\'al]{Feher}
{\sc \au{Feh\'er, T.}, \au{Smith, H.~M.}, \au{F\"ul\"op, T.} \& \au{G\'al, K.}}
  \yr{2011}  \at{Simulation of runaway electron generation during plasma
  shutdown by impurity injection in {ITER}}.  \jt{Plasma Physics and Controlled
  Fusion}  \bvol{53}~(3),  \pg{035014}.

\bibitem[G\'al {\em et~al.\/}(2008)G\'al, Feh\'er, Smith, F\"ul\"op \&
  Helander]{gal}
{\sc \au{G\'al, K.}, \au{Feh\'er, T.}, \au{Smith, H.~M.}, \au{F\"ul\"op, T.} \&
  \au{Helander, P.}} \yr{2008}  \at{Runaway electron generation during plasma
  shutdown by killer pellet injection}.  \jt{Plasma Phys. and Controlled
  Fusion}  \bvol{50},  \pg{055006}.

\bibitem[Greenwald {\em et~al.\/}(2018)Greenwald, Whyte, Bonoli, Hartwig, Irby,
  LaBombard, Marmar, Minervini, Takayasu, Terry, Vieira, White, Wukitch,
  Brunner, Mumgaard \& Sorbom]{SPARC}
{\sc \au{Greenwald, M.}, \au{Whyte, D.}, \au{Bonoli, P.}, \au{Hartwig, Z.},
  \au{Irby, J.}, \au{LaBombard, B.}, \au{Marmar, E.}, \au{Minervini, J.},
  \au{Takayasu, M.}, \au{Terry, J.}, \au{Vieira, R.}, \au{White, A.},
  \au{Wukitch, S.}, \au{Brunner, D.}, \au{Mumgaard, R.} \& \au{Sorbom, B.}}
  \yr{2018} {The high-field path to practical fusion energy}.

\bibitem[Helander \& Sigmar(2005)]{helander}
{\sc \au{Helander, P.} \& \au{Sigmar, D.}} \yr{2005} {\em Collisional Transport
  in Magnetized Plasmas\/}.  \publ{Cambridge University Press}.

\bibitem[Helander {\em et~al.\/}(2004)Helander, Smith, F\"ul\"op \&
  Eriksson]{helanderhottail}
{\sc \au{Helander, P.}, \au{Smith, H.~M.}, \au{F\"ul\"op, T.} \& \au{Eriksson,
  L.~G.}} \yr{2004}  \at{Electron kinetics in a cooling plasma}.  \jt{Physics
  of Plasmas}  \bvol{11},  \pg{5704}.

\bibitem[Hesslow {\em et~al.\/}(2019{\natexlab{{\em a\/}}})Hesslow,
  Embr{\'{e}}us, Vallhagen \& F\"ul\"op]{Hesslow_2019}
{\sc \au{Hesslow, L.}, \au{Embr{\'{e}}us, O.}, \au{Vallhagen, O.} \&
  \au{F\"ul\"op, T.}} \yr{2019{\natexlab{{\em a\/}}}}  \at{Influence of massive
  material injection on avalanche runaway generation during tokamak
  disruptions}.  \jt{Nuclear Fusion}  \bvol{59}~(8),  \pg{084004}.

\bibitem[Hesslow {\em et~al.\/}(2019{\natexlab{{\em b\/}}})Hesslow, Unnerfelt,
  Vallhagen, Embreus, Hoppe, Papp \& F\"ul\"op]{Hesslowupcoming}
{\sc \au{Hesslow, L.}, \au{Unnerfelt, L.}, \au{Vallhagen, O.}, \au{Embreus,
  O.}, \au{Hoppe, M.}, \au{Papp, G.} \& \au{F\"ul\"op, T.}}
  \yr{2019{\natexlab{{\em b\/}}}}  \at{Evaluation of the {Dreicer} runaway
  generation rate in the presence of high-{Z} impurities using a neural
  network}.  \jt{Journal of Plasma Physics}  \bvol{85}~(6),  \pg{475850601}.

\bibitem[Hollmann {\em et~al.\/}(2013)Hollmann, Austin, Boedo, Brooks, Commaux,
  Eidietis, Humphreys, Izzo, James, Jernigan, Loarte, Martin-Solis, Moyer,
  Mu{\~{n}}oz-Burgos, Parks, Rudakov, Strait, Tsui, Zeeland, Wesley \&
  Yu]{Hollmann_2013}
{\sc \au{Hollmann, E.}, \au{Austin, M.}, \au{Boedo, J.}, \au{Brooks, N.},
  \au{Commaux, N.}, \au{Eidietis, N.}, \au{Humphreys, D.}, \au{Izzo, V.},
  \au{James, A.}, \au{Jernigan, T.}, \au{Loarte, A.}, \au{Martin-Solis, J.},
  \au{Moyer, R.}, \au{Mu{\~{n}}oz-Burgos, J.}, \au{Parks, P.}, \au{Rudakov,
  D.}, \au{Strait, E.}, \au{Tsui, C.}, \au{Zeeland, M.~V.}, \au{Wesley, J.} \&
  \au{Yu, J.}} \yr{2013}  \at{Control and dissipation of runaway electron beams
  created during rapid shutdown experiments in {DIII}-{D}}.  \jt{Nuclear
  Fusion}  \bvol{53}~(8),  \pg{083004}.

\bibitem[Hollmann {\em et~al.\/}(2015)Hollmann, Aleynikov, F\"ul\"op,
  Humphreys, Izzo, Lehnen, Lukash, Papp, Pautasso, Saint-Laurent \&
  Snipes]{HollmannDMS}
{\sc \au{Hollmann, E.~M.}, \au{Aleynikov, P.~B.}, \au{F\"ul\"op, T.},
  \au{Humphreys, D.~A.}, \au{Izzo, V.~A.}, \au{Lehnen, M.}, \au{Lukash, V.~E.},
  \au{Papp, G.}, \au{Pautasso, G.}, \au{Saint-Laurent, F.} \& \au{Snipes,
  J.~A.}} \yr{2015}  \at{Status of research toward the {ITER} disruption
  mitigation system}.  \jt{Physics of Plasmas}  \bvol{22}~(2),  \pg{021802}.

\bibitem[Izzo {\em et~al.\/}(2011)Izzo, Hollmann, James, Yu, Humphreys, Lao,
  Parks, Sieck, Wesley, Granetz, Olynyk \& Whyte]{Izzo_2011}
{\sc \au{Izzo, V.}, \au{Hollmann, E.}, \au{James, A.}, \au{Yu, J.},
  \au{Humphreys, D.}, \au{Lao, L.}, \au{Parks, P.}, \au{Sieck, P.}, \au{Wesley,
  J.}, \au{Granetz, R.}, \au{Olynyk, G.} \& \au{Whyte, D.}} \yr{2011}
  \at{Runaway electron confinement modelling for rapid shutdown scenarios in
  {DIII}-{D}, {A}lcator {C-M}od and {ITER}}.  \jt{Nuclear Fusion}
  \bvol{51}~(6),  \pg{063032}.

\bibitem[Izzo {\em et~al.\/}(2012)Izzo, Humphreys \& Kornbluth]{Izzo_2012}
{\sc \au{Izzo, V.~A.}, \au{Humphreys, D.~A.} \& \au{Kornbluth, M.}} \yr{2012}
  \at{Analysis of shot-to-shot variability in post-disruption runaway electron
  currents for diverted {DIII}-{D} discharges}.  \jt{Plasma Physics and
  Controlled Fusion}  \bvol{54}~(9),  \pg{095002}.

\bibitem[Jayakumar {\em et~al.\/}(1993)Jayakumar, Fleischmann \&
  Zweben]{jayakumar1993}
{\sc \au{Jayakumar, R.}, \au{Fleischmann, H.} \& \au{Zweben, S.}} \yr{1993}
  \at{Collisional avalanche exponentiation of runaway electrons in electrified
  plasmas}.  \jt{Physics Letters A}  \bvol{172},  \pg{447 -- 451}.

\bibitem[Landreman {\em et~al.\/}(2014)Landreman, Stahl \&
  F\"ul\"op]{CODEPaper2014}
{\sc \au{Landreman, M.}, \au{Stahl, A.} \& \au{F\"ul\"op, T.}} \yr{2014}
  \at{Numerical calculation of the runaway electron distribution function and
  associated synchrotron emission}.  \jt{Computer Physics Communications}
  \bvol{185}~(3),  \pg{847}.

\bibitem[Martin \& Shaw(2019)]{subatomic}
{\sc \au{Martin, B.~R.} \& \au{Shaw, G.}} \yr{2019} {\em Nuclear and Particle
  Physics : An Introduction\/}.  \publ{John Wiley \& Sons, Incorporated}.

\bibitem[Mart\'{\i}n-Sol\'{\i}s {\em et~al.\/}(2015)Mart\'{\i}n-Sol\'{\i}s,
  Loarte \& Lehnen]{martinsolis1}
{\sc \au{Mart\'{\i}n-Sol\'{\i}s, J.~R.}, \au{Loarte, A.} \& \au{Lehnen, M.}}
  \yr{2015}  \at{Runaway electron dynamics in tokamak plasmas with high
  impurity content}.  \jt{Physics of Plasmas}  \bvol{22},  \pg{092512}.

\bibitem[Mart{\'{\i}}n-Sol{\'{\i}}s {\em
  et~al.\/}(2017)Mart{\'{\i}}n-Sol{\'{\i}}s, Loarte \& Lehnen]{MartinSolis2017}
{\sc \au{Mart{\'{\i}}n-Sol{\'{\i}}s, J.~R.}, \au{Loarte, A.} \& \au{Lehnen,
  M.}} \yr{2017}  \at{Formation and termination of runaway beams in {ITER}
  disruptions}.  \jt{Nuclear Fusion}  \bvol{57}~(6),  \pg{066025}.

\bibitem[Papp {\em et~al.\/}(2013)Papp, F\"ul\"op, Feh{\'{e}}r, de~Vries,
  Riccardo, Reux, Lehnen, Kiptily, Plyusnin \& and]{Papp_2013}
{\sc \au{Papp, G.}, \au{F\"ul\"op, T.}, \au{Feh{\'{e}}r, T.}, \au{de~Vries,
  P.}, \au{Riccardo, V.}, \au{Reux, C.}, \au{Lehnen, M.}, \au{Kiptily, V.},
  \au{Plyusnin, V.} \& \au{and, B.~A.}} \yr{2013}  \at{The effect of
  {ITER}-like wall on runaway electron generation in {JET}}.  \jt{Nuclear
  Fusion}  \bvol{53}~(12),  \pg{123017}.

\bibitem[Reux {\em et~al.\/}(2015)Reux, Plyusnin, Alper, Alves, Bazylev,
  Belonohy, Boboc, Brezinsek, Coffey, Decker, Drewelow, Devaux, de~Vries, Fil,
  Gerasimov, Giacomelli, Jachmich, Khilkevitch, Kiptily, Koslowski, Kruezi,
  Lehnen, Lupelli, Lomas, Manzanares, Aguilera, Matthews, Mlyn{\'a}{\v{r}},
  Nardon, Nilsson, von Thun, Riccardo, Saint-Laurent, Shevelev, Sips, Sozzi \&
  contributors]{Reux2015}
{\sc \au{Reux, C.}, \au{Plyusnin, V.}, \au{Alper, B.}, \au{Alves, D.},
  \au{Bazylev, B.}, \au{Belonohy, E.}, \au{Boboc, A.}, \au{Brezinsek, S.},
  \au{Coffey, I.}, \au{Decker, J.}, \au{Drewelow, P.}, \au{Devaux, S.},
  \au{de~Vries, P.}, \au{Fil, A.}, \au{Gerasimov, S.}, \au{Giacomelli, L.},
  \au{Jachmich, S.}, \au{Khilkevitch, E.}, \au{Kiptily, V.}, \au{Koslowski,
  R.}, \au{Kruezi, U.}, \au{Lehnen, M.}, \au{Lupelli, I.}, \au{Lomas, P.},
  \au{Manzanares, A.}, \au{Aguilera, A. M.~D.}, \au{Matthews, G.},
  \au{Mlyn{\'a}{\v{r}}, J.}, \au{Nardon, E.}, \au{Nilsson, E.}, \au{von Thun,
  C.~P.}, \au{Riccardo, V.}, \au{Saint-Laurent, F.}, \au{Shevelev, A.},
  \au{Sips, G.}, \au{Sozzi, C.} \& \au{contributors, J.}} \yr{2015}
  \at{Runaway electron beam generation and mitigation during disruptions at
  {JET-ILW}}.  \jt{Nuclear Fusion}  \bvol{55}~(9),  \pg{093013}.

\bibitem[Rosenbluth \& Putvinski(1997)]{RosenbluthPutvinski1997}
{\sc \au{Rosenbluth, M.} \& \au{Putvinski, S.}} \yr{1997}  \at{Theory for
  avalanche of runaway electrons in tokamaks}.  \jt{Nuclear Fusion}  \bvol{37},
   \pg{1355--1362}.

\bibitem[Smith {\em et~al.\/}(2006)Smith, Helander, Eriksson, Anderson, Lisak
  \& Andersson]{Smith2006GO}
{\sc \au{Smith, H.}, \au{Helander, P.}, \au{Eriksson, L.-G.}, \au{Anderson,
  D.}, \au{Lisak, M.} \& \au{Andersson, F.}} \yr{2006}  \at{Runaway electrons
  and the evolution of the plasma current in tokamak disruptions}.  \jt{Physics
  of Plasmas}  \bvol{13}~(10),  \pg{102502}.

\bibitem[Wilson(1925)]{wilson1925}
{\sc \au{Wilson, C. T.~R.}} \yr{1925}  \at{The acceleration of
  $\beta$-particles in strong electric fields such as those of thunderclouds}.
  \jt{Mathematical Proceedings of the Cambridge Philosophical Society}
  \bvol{22},  \pg{534}.

\end{thebibliography}
\end{document}